# Graphene FET Process and Analysis Optimization in 200 mm Pilot Line Environment

*Anton Murros[1,§], Miika Soikkeli[1,*,§], Anni Virta[1], Arantxa Maestre[2,3], Leire Morillo[2], Alba Centeno[2], Amaia Zurutuza[2], Olli-Pekka Kilpi[1].*

[1]VTT Technical Research Centre of Finland Ltd, P.O. Box 1000, FI-02044 VTT, Espoo, Finland

[2]Graphenea S.A., Paseo Mikeletegi 83, 20009-San Sebastian, Spain

[3]ASM International, Kapeldreef 75, 3001 Leuven, Belgium

[§]These authors contributed equally to this work.

The maturity of the chemical vapor deposition graphene-based device processing has increased from chip level demonstrations to wafer-scale fabrication in the past few years. Due to this wafer-scale, electrical characterization and analysis of the fabricated devices has become increasingly important to enable extraction of multiple parameters with minimal number of measurements for the quality control purposes critical for industrial uptake of 2D materials-based devices. As a crucial step, we demonstrate optimization of complementary metal-oxide semiconductor (CMOS) back-end-of-line (BEOL) compatible graphene field-effect transistor (GFET) fabrication and analysis including the gate stack, bottom contact, graphene patterning and encapsulation process steps. The analysis methods include atomic force microscopy, scanning electron microscopy and most importantly electrical characterization. The electrical characterization focuses on comparing different test structures and extraction methods for mobility, contact resistance, IV-curve hysteresis and doping parameters. The comparison shows that the selected measurement test structures and analysis methods can have a large impact on the extracted values and should thus be considered when comparing data sets between different sources. The analysis shows that the optimized process offers high device yield of 98 % with good doping uniformity, contact resistance and mobility as well as low IV-curve hysteresis values on 200 mm wafers.

# 1. Introduction

Graphene, first isolated in 2004 using mechanical exfoliation, marked the beginning of a new era in two-dimensional materials research due to its exceptional electrical and mechanical properties.[1] Currently, after over 20 years of research, the maturity of graphene-based technologies is transitioning from the chip-level fabrication and analysis of single devices to wafer-scale fabrication and statistical analysis of numerous devices within a wafer and between different wafers. Wafer-scale chemical vapor deposition (CVD) graphene-based device fabrication has been demonstrated already on 200 mm and 300 mm wafer platforms for sensing and photonics application areas thanks to the advancements in the growth, transfer, and processing of the 2D materials made in recent years.[2–9] These advancements are critical for the industrial uptake of the 2D materials in clearing the barriers for the integration into semiconductor fabrication lines.[10]

It is critical to ensure that the CVD graphene-based sensor fabrication processes are compatible with CMOS back end of line (BEOL) as this can bring the applications to higher maturity level for example for broadband image detectors[11], gas sensing[12] and biosensing[4]. The relatively advanced technological level achieved in these studies shows significant progress in the monolithic integration of 2D material sensors into advanced CMOS systems. The possibilities and challenges of graphene integration into CMOS back end of line (BEOL) have been discussed by Neumaier et al.[10] However, many of the presented integration flows only offer limited functionality of the devices in CMOS BEOL integration for sensing applications due to the lacking gates[11], global Si back gates[13] and liquid gates[4] instead of local back gates. The local back gates offer clear benefits for quality control and sensitivity tuning of the devices.[2] These are especially beneficial for gas sensors[14] and photodetectors[2], while biosensors[4,15] typically favor the usage of liquid gates instead of local back gates in actual biosensing scenarios but can also gain increased sensitivity with dual gate[16,17] sensing configurations.

Large scale adaptation of GFETs requires quality control protocols with detailed extraction of multiple parameters for detailed quality control of the fabricated GFETs,[18,19] however measurements and analysis of wafer-scale GFETs is often conducted to extract only single or few electrical parameters such as mobility[13], contact resistance[20–22], doping[13] and IV-curve hysteresis[23], which might be adequate for a specific application but only provides limited information on the quality of the devices. [17,18] The measurement conditions in which the device characterization is performed should also be carefully considered since the ambient measurement conditions heavily affect the obtained parameters and thus the device parameters should be obtained either in vacuum[24,25] or with passivated devices[13].

Electrical characterization of graphene-based devices is typically done by using for example transfer length measurement (TLM) with and without gating[26,27], Hall bar[28], Van der Pauw[29,30], cross-bridge Kelvin (CBK)[31,32], contact-end-resistance (CER) [25,32–34], GFET[13] and capacitor[35] test structures. These methods are targeted to extract material and device parameters, such as contact resistance, doping and mobility. Different structures have pros and cons which affect reliability of the methods. Gated TLM (gTLM) structures offer the possibility to normalize the transfer curves with respect to Dirac peak position, leading to more accurate extraction of parameters at specific carrier concentrations. Hall bar and Van der Pauw structures can be utilized to measure sheet resistance of the graphene devices directly with 4-point measurement configuration for mobility extraction.[28–30] Mobility calculation from the extracted or measured sheet resistance values can then be done for example by using transconductance or Drude methods. As an alternative contact resistance extraction method for graphene devices, CBK and CER structures have been utilized.[34,36] Individual GFETs can be directly analyzed by using the direct transconductance method (DTM) for mobility or by a fitting method (FTM)[37] for mobility and contact resistance values.[26] For accurate estimation of the mobility and doping values, through field-effect measurement, capacitor test structures need be utilized to measure gate capacitance instead of using theoretical values for the dielectric constant especially for the thinner gate oxides where quantum capacitance ($C_q$) of graphene influences the gate capacitance.[35]

In this work, we demonstrate optimization of wafer-scale GFET fabrication and device characterization towards high device yield and uniformity. The focus is on the GFET processing including gate, contact, graphene and encapsulation modules compatible with CMOS BEOL integration. Measurement and analysis of the fabricated GFETs are done on wafer-scale with a specific set of test structures with varying device dimensions distributed evenly across the wafer within process control monitor (PCM) areas. The objective is to extract mobility, contact resistance, IV-hysteresis and doping of the devices with a minimal number of measurements to develop a protocol for quality control for the GFET fabrication in a pilot line environment. The work done here further enables the high yield and uniformity BEOL CMOS integration and quality control of GFETs for applications such as biosensors, gas sensors and infrared cameras.

## 2. Materials and methods

**GFET processing**

We present optimization of the GFET fabrication processes including gate, contact, graphene and encapsulation modules. Bottom contact process has been selected to minimize the processing steps after the graphene transfer step. The benefit of this approach is to reduce the residues and simplify the cleaning of the graphene surface. Overall, 6 different wafer runs have been fabricated to optimize these modules (Runs 1–6). In addition, a final run with 4 wafers has been fabricated to study the difference between one- and two-layer resist lithography for the graphene patterning step (Run 7).

The cross-section and process parameters of the GFETs are shown in Figures 1a and 1b, respectively. 150 mm Si wafers with 1000 nm thermal $SiO_2$ were used as the starting substrates for Runs 1–4. Runs 5–6 were done on 200 mm Si wafers with 300 nm thermal $SiO_2$ for Run 5 and 300 nm LPCVD $Si_3N_4$ for Run 6. The first process step includes the local back gate 60 nm Al / 15 nm NbN deposition and dry etching for Run 1 and 3 nm Ti / 30 nm Au metal deposition by evaporation and lift-off for the following 5 runs (Runs 2–6). The lithography process was optimized between Runs 2–4 to ensure good lift-off profile. The gate dielectric of 50 nm $Al_2O_3$ for Runs 1-6 was deposited with ALD. The bottom contact 3 nm Ti / 30 nm Au deposition was done by a lift-off process and graphene transfer was done with semi-dry transfer process by Graphenea. The graphene was patterned with photolithography and $O_2$ plasma dry etching. The devices were encapsulated by first evaporating a 1 nm Al seed layer that was oxidized, followed by 50 nm ALD $Al_2O_3$ and an additional 100 nm PECVD $Si_3N_4$ layer. Finally, the encapsulation was opened by dry etching the $Si_3N_4$ and wet etching the $Al_2O_3$ from the sensor devices but not from the quality control devices used to extract the electrical parameters. The process differences are summarized in Figure 1b.

The process flow for Run 7 for the final 4 wafers was slightly modified from Run 6 process to study the impact of one- and two-layer resist lithography on the amount of resist residues on the patterned graphene channel and on doping uniformity. In addition, the gate dielectric was changed to 17 nm of $Al_2O_3$ and 3 nm of $HfO_2$. The target of these wafers is to develop a biosensing compatible process flow where a highly selective channel opening etch is possible and the surface cleanliness of graphene is improved. Graphene patterning was done for 2 wafers with one-layer approach with AZ5214 and for 2 wafers with two-layer approach similar to the solution by Gao et al.[38] with polydimethylglutarimide (PMGI) and AZ5214. These devices were not encapsulated in this study.

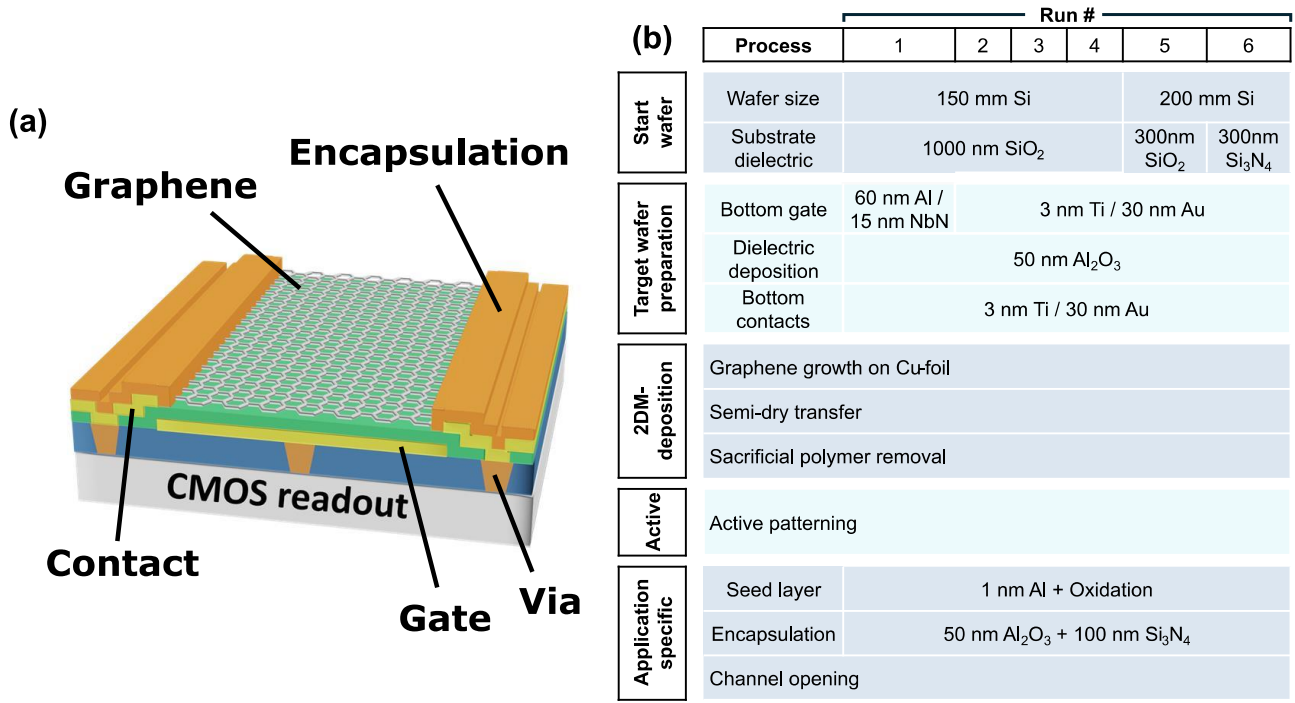

**Figure 1** a) A schematic of the basic modules included in the GFET fabrication process on a CMOS readout wafer. The modules include back gate deposition, bottom contact deposition, graphene transfer, patterning and encapsulation options. b) The process flow of the GFET fabrication for Runs 1–6.

**Wafer-scale device characterization**

The performance of the GFETs was characterized by using a semiconductor parameter analyzer, impedance analyzer and an automated probe station to map out the wafer-scale statistics for each wafer. An example of the set of test structures used is shown in Figure 2. The test structures include GFETs (Figure 2a), gTLMs (Figure 2b), metal-insulator-metal (MIM) and metal-insulator-semiconductor (MIS) capacitors (Figure 2c), and gated CBK (gCBK) resistors (Figure 2d). The resistance of the GFET and gTLM structures is measured for a series of channel lengths (5–50 μm) and widths (5–50 μm). These are analyzed using the gated TLM method to extract both mobility and contact resistance. We will distinguish between the test structures and analysis method by referring to the prior as gTLMs, and the latter as the gated TLM method. We also compare the use of DTM and FTM to extract mobility and contact resistance values. The GFETs and gTLMs include additional leads which enable 4-point Kelvin measurements to remove parasitic resistance. Direct contact resistance measurements are performed by using the gCBK structures. Capacitor test structures are used to estimate the gate capacitance value of the devices for more precise evaluation of parameters such as mobility and doping. Doping and IV-curve hysteresis values are extracted directly from the measured GFET and gTLM channel IV-curves.

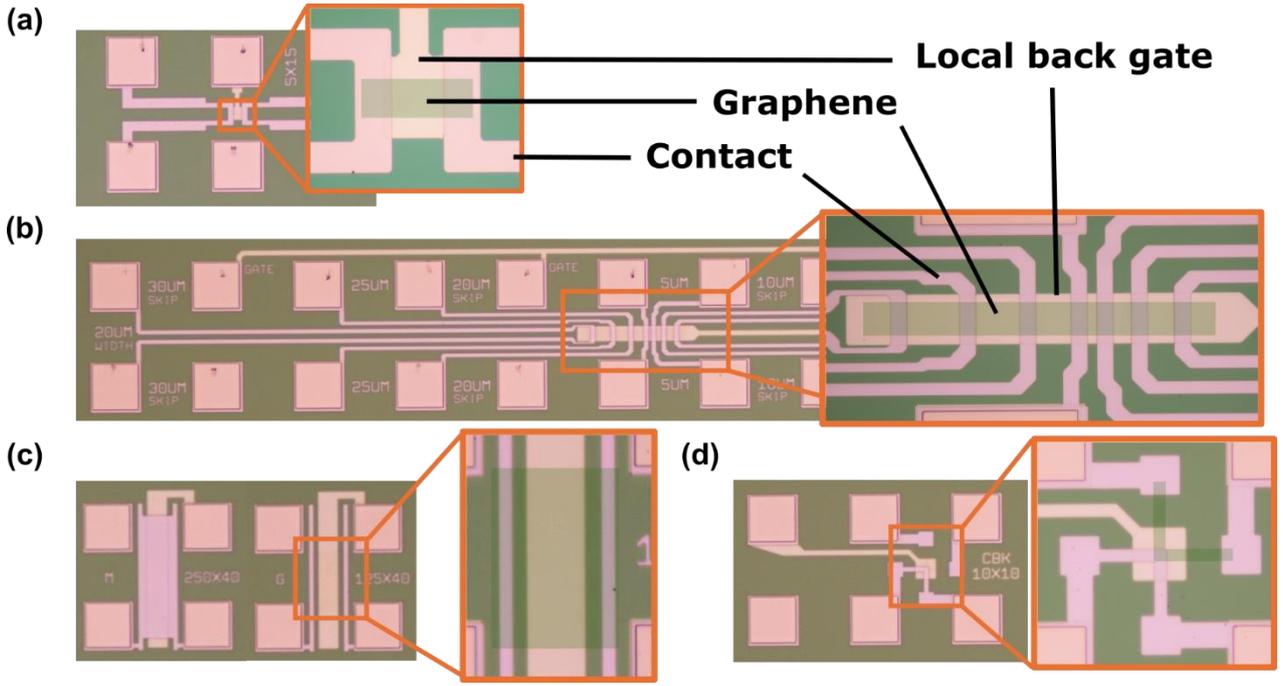

**Figure 2.** Electrical test structures with a) a GFETs, b) a gTLM structure, c) MIM (left) and MIS (right) capacitors and d) gCBKs. Inset in a)–d) are larger magnification images with the graphene channel pseudo-coloured.

The GFETs and gTLMs were analyzed in a similar manner. A constant drain-source bias of 100mV is applied and the total resistance $R_{total}$ is measured as a function of gate voltage $V_{GS}$. The transfer curves are normalized to the Dirac peak voltage ($V_{Dirac}$). Linear fitting is then used to extract sheet and contact resistance as a function of gate voltage overdrive ($V_{GSO} = V_{GS} - V_{Dirac}$). The extraction method for the contact and sheet resistances at some gate voltage overdrive is shown in Figure 3. $R_{total}$ consists of channel resistance ($R_{channel}$) from graphene and contact resistance ($R_c$) from the graphene/metal junction. It can be expressed as:

$$R_{total} = 2R_c + R_{channel} = 2R_c + \frac{R_{sh}}{W}L, \qquad (1)$$

where $R_{sh}$, L and W are the sheet resistance, length and width of graphene. By linear fitting $R_{total}$ of graphene channels with width W and different L at each $V_{GSO}$ point, $R_{sh}$ and $R_c$ can be extracted from the slope and intercept, respectively, and the $V_{GSO}$ dependent $R_{sh}$ and $R_c$ curves can be obtained.

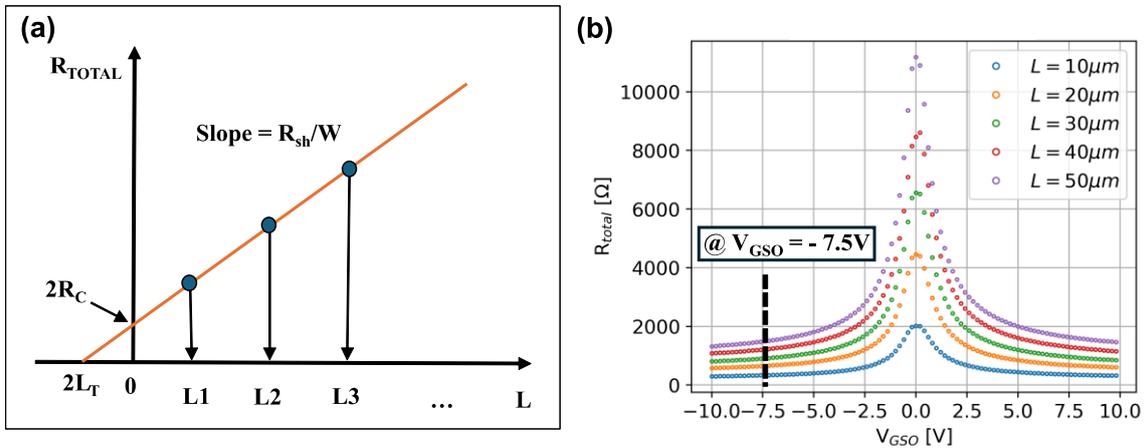

**Figure 3.** a) Determination of contact and sheet resistance by linear fitting the measured resistance values of different channel lengths at some gate voltage overdrive. b) Measured GFET transfer curves with W = 20 μm and L = 10–50 μm.

Mobility as a function of gate voltage is then calculated by using the extracted sheet resistance according to the Drude model[39]:

$$\sigma_{sh} = \frac{1}{R_{sh}} = \mu e n, \tag{2}$$

where e is the elementary charge, $1.6 \times 10^{-19}$ C, and n is the gate induced charge carrier or doping concentration, which can be estimated by:

$$n = \frac{C_g V_{GSO}}{e}, \tag{3}$$

where $C_g$ is gate capacitance per unit area. This value is measured using an impedance analyser from the capacitor test structures to get more accurate mobility values (see Supplementary Information Section 1). We measure $C_g \approx 1.45 \times 10^{3}$ F/m$^{-2}$ for the gate dielectric in Runs 1–6.

The gated TLM method mobility, $\mu^{TLM}$, can thus be written as:

$$\mu^{TLM} = \frac{\frac{1}{R_{sh}}}{V_{GSO} C_g}. \tag{4}$$

It should be noted that the Drude model is not directly applicable for small $V_{GSO}$[39]. The hole and electron mobilities are extracted at carrier concentrations of $-1 \times 10^{12}$ cm$^{-2}$ and $1 \times 10^{12}$ cm$^{-2}$, respectively.

The mobility of the graphene channel can also be directly calculated using the Drude model on individual channels, where the sheet conductance can be estimated using the dimensions of the graphene channel[41]:

$$\mu^{Drude} = \frac{1}{R_{sh}} \frac{1}{V_{GSO} C_g} = \frac{L}{R_{channel} W} \frac{1}{V_{GSO} C_g} \approx \frac{I_{ds} L}{V_{ds} W} \frac{1}{V_{GSO} C_g}, \tag{5}$$

where $R_{sh} = R_{channel} W/L$ and $R_{channel} \approx V_{ds}/I_{ds}$. This method neglects the contribution of $R_c$ and is only applicable when $R_c$ is small.

Alternatively, the DTM can be used to estimate mobility.[26] In this case mobility is extracted from the gate voltage dependent transconductance by using the following equation:

$$\mu^{DTM} = g_m \frac{L}{W V_{ds} C_g}, \tag{6}$$

where $g_m = \partial I_{ds}/\partial V_{GSO}$ is transconductance. The DTM is applied to individual GFET transfer curves. Typically, hole and electron mobilities are taken from the maxima of the hole and electron conduction branches, respectively. Here we define the hole $\mu^{DTM}$ at a carrier concentration of $-1 \times 10^{12}$ cm$^{-2}$ to be comparable with the other methods.

Equation (2) can be modified to consider the total charge carrier concentration, defined as $n_{tot} = \sqrt{n_0^2 + n^2}$, where $n_0$ is the residual charge carrier concentration at minimum conductivity (Dirac point), which is ideally zero for a disorder-free graphene.[37] Using this definition we can directly model $R_{total}$ from Equation (1):

$$R_{total} = 2R_c + \frac{R_{sh}}{W} L = 2R_c + \frac{\frac{1}{\mu^{FTM} e n_{tot}}}{W} L = 2R_c + \frac{L}{W} \frac{1}{\mu^{FTM} e n_{tot}}. \tag{7}$$

Equation (7) is a commonly used model for mobility and contact resistance extraction[37]. Here we have an additional factor of 2 with the $R_c$ term to be consistent with Equation (1). Contact resistance and mobility can be extracted by fitting Equation (7) against measured transfer curves. We define $\mu^{FTM}$ to differentiate between the result of directly applying Equation (5). See Supplementary Information Section 2 for more details.

Finally, we also compare the mobility of the graphene measured using the Hall effect on square Van der Pauw test structures by using Ecopia HMS-500.

IV-curve hysteresis has been measured by sweeping the gate voltage back and forth. The difference between the Dirac peak voltages normalized by dividing with the total voltage range is used as the hysteresis value according to the following equation:

$$Hysteresis = \frac{(V_{Dirac,f} - V_{Dirac,b})}{V_{range}}. \tag{8}$$

gCBK structures have been utilized as an alternative approach to measure contact resistance directly to compare with the values obtained by using the gated TLM method. The measurement principle is presented in Supplementary Information Section 3 (see Supplementary Information Figure S5). A current is applied between contacts 1 and 2 while the voltage drop over the contact is measured between contacts 3 and 4. Contact resistance can then be obtained according to the following equation:

$$R_c = \frac{V_{34}}{I}, \tag{9}$$

where $V_{34}$ is the voltage drop between contact 3 and 4 and $I$ is the current between points 1 and 2. Here we normalize the contact resistance for the graphene devices by the width of the graphene instead of the contact area as typically done with gCBK structures.[36] $R_c$ extracted using the other methods is also normalized in this manner to define specific contact resistance:

$$\rho_c = R_c W. \tag{10}$$

Finally, yield of the devices has been defined with on/off ratio, maximum channel resistance and gate leakage current parameters. On/off ratio should be higher than 5, channel resistance smaller than 1 MΩ and leakage current smaller than 10 nA for the device to be accepted as a working device.

**AFM and SEM characterization**

Atomic force microscopy (AFM) and scanning electron microscopy (SEM) were used to analyze the topographical results of the process optimization efforts to reduce both the contact and back gate roughness, step heights and edge discontinuities. Here edge discontinuity refers to additional material at the edges of metal features. SEM was used to qualitatively analyze layer step heights and edge discontinuities, and AFM was used to quantitatively measure surface roughness, layer step heights and edge discontinuities. In addition, AFM was used to analyze the amount of resist residues on the graphene channel by measuring the total thickness of the channel after the graphene patterning step.

# 3. Results and discussion

**Back gate and bottom contact optimization**

**SEM and AFM characterization**

The contact and back gate roughness, step heights and discontinuities were characterized by AFM and SEM. In Run 1, the back gate roughness and step height were 1.9 nm and 77 nm, respectively (Figure 4a). The back gate patterning method was changed from etching to lift-off and material from Al/NbN to Ti/Au after Run 1. This resulted in lower back gate metal RMS roughness of 0.9 nm and step height of 33 nm (Figure 4b). The process change to lift-off resulted in an edge discontinuity presenting as additional edge height of 110 nm in Run 2. In Run 3 (not shown) this was reduced to 60 nm, and to 0 nm in Run 4 (Figure 4c). The surface roughness was further optimized in Runs 3–4 finally resulting in a back gate RMS roughness and step height of 0.7 nm and 33 nm, respectively. Runs 5–7 had similar roughness and step height values as obtained in Run 4 without edge discontinuity issues.

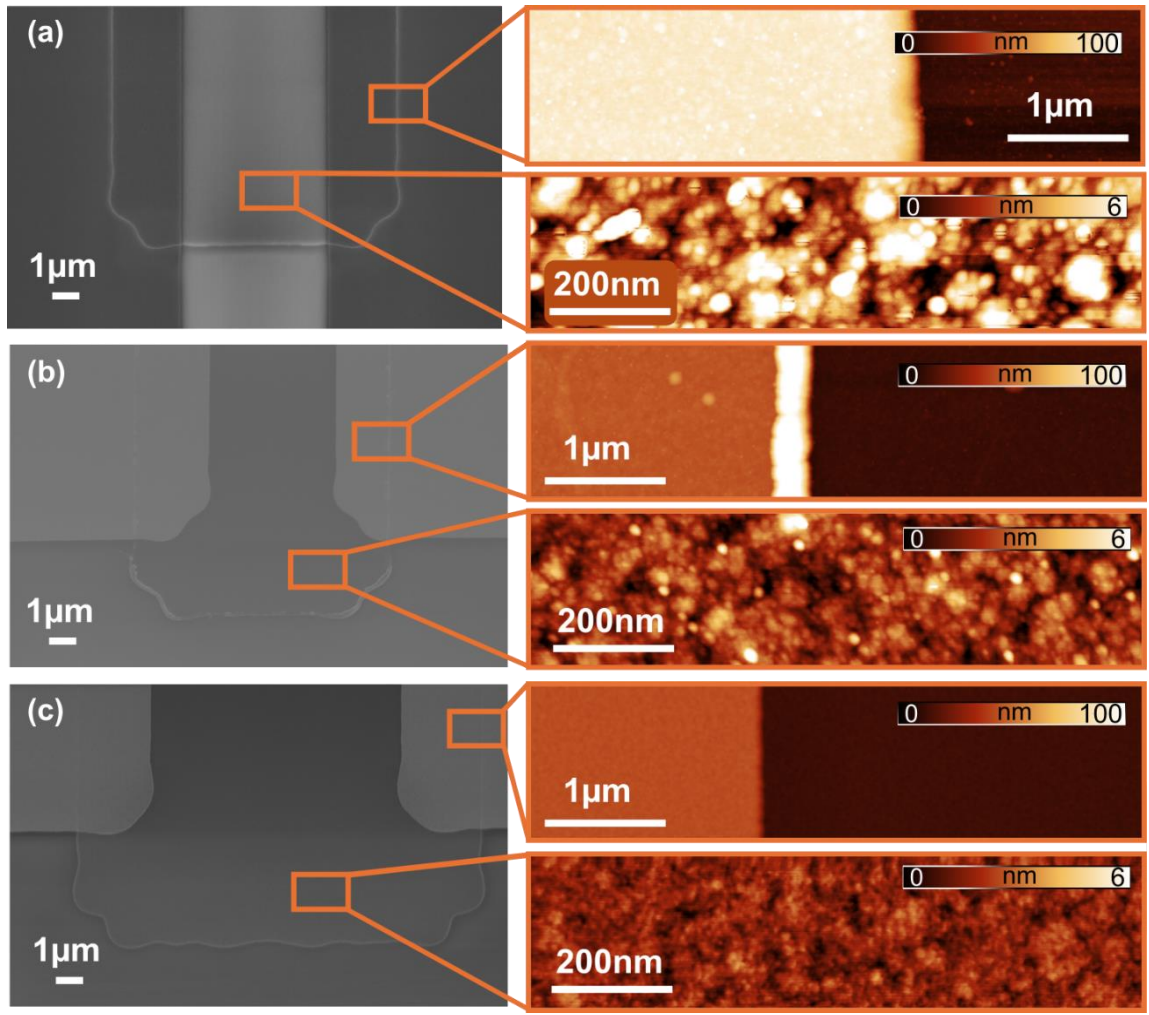

**Figure 4.** SEM and AFM images of a local back gate and bottom contacts. The inset AFM images depict the gate metal edge (upper) and back gate metal roughness (lower). a) Run 1: 77 nm back gate step height and roughness $R_q$ = 1.9 nm. b) Run 2: 33 nm back gate step height, roughness $R_q$ = 0.9 nm and 110 nm edge discontinuity. c) Run 4: 33 nm back gate step height, roughness $R_q$ = 0.7 nm and no edge discontinuity.

## Wafer-scale electrical characterization – GFET test structures with gated TLM method

Specific contact resistance, mobility, doping and yield are extracted for the wafers in Runs 1–6 using Equations (1), (3) and (4) according to the criteria described earlier. For each wafer a series of GFETs were measured ranging from L = 5–50 μm and W = 5–50 μm, with a back gate voltage range of -10–10 V. Individual GFET channels (Figure 1a) were measured for each width and length, and gated TLM fitting was performed on a set of GFETs with the same width. The wafer-scale statistics for the extracted parameters are shown in Figure 5. For Runs 1–6 the number of GFETs and TLM channels measured for each wafer is 768, 2304, 1536, 1536, 552 and 8640, respectively. The number of devices measured for Run 5 is smaller due to PCM area restrictions on the wafer. The average width-normalized specific contact resistance ($\rho_c$) values at a carrier concentration of $7.5 \times 10^{12}$ cm$^{-2}$ (Figure 5a) are 2250 (SD = 1420) Ωμm, 830 (SD = 1390) Ωμm, 320 (SD = 300) Ωμm, 270 (SD = 470) Ωμm, 280 (SD = 460) Ωμm, and 600 (SD = 350) Ωμm for Runs 1–6, respectively. The data includes also the non-physical negative $\rho_c$ values which are quite common in large data sets with fitting-based extraction relying on data from several graphene channels due to the variations between the channels, such as fabrication defects including for example bilayer regions and holes as well as different graphene sheet resistance values under the metal and channel areas. [25,32–34,42,43]

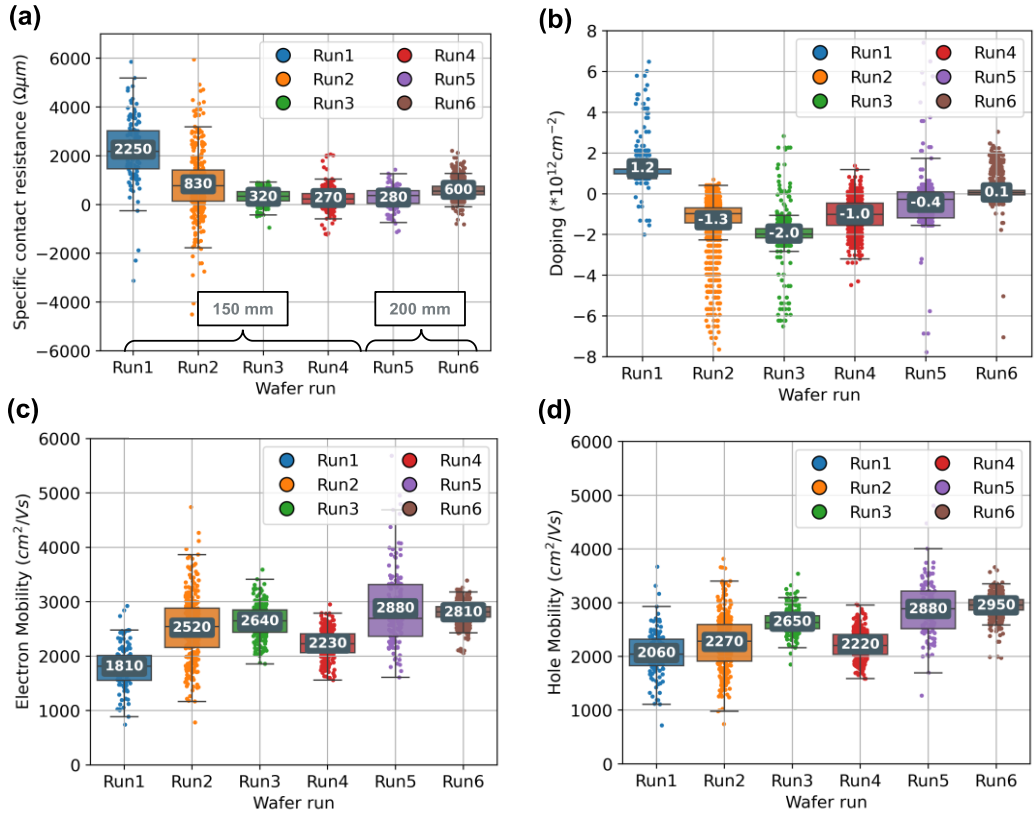

**Figure 5.** Statistical analysis of process Runs 1–6 with electrical measurements for a) specific contact resistance, b) doping, c) electron mobility and d) hole mobility. Inset values are average values of each wafer.

The run-by-run improvement in $\rho_c$, shown in Figure 5, demonstrate the importance in the reduction of the step heights and back gate roughness from Runs 1–2 and the edge discontinuities from Runs 2–4. Similar values for $\rho_c$ were achieved in Run 5 when the process was scaled from 150 mm to 200 mm wafers. Average $\rho_c$ slightly increased from Run 5 to Run 6, which was partially attributed to an increased number of defects in the 5 μm wide channels, but there were also less non-physical negative results which also contributed to the increase of the average value. For hole mobility ($\mu_h^{TLM}$) (Figure 5d) we also observe an upward trend starting from the average value of 2060 (SD = 440) cm$^2$/Vs in Run 1 and ending in average value of 2950 (SD = 170) cm$^2$/Vs in Run 6. We believe this increase is likely due to the decrease in the roughness on the gate stack area. The average doping values (Figure 5b) range from -2.0 to 1.2 (SD ≈ 1) × 10$^{12}$ cm$^{-2}$ for Runs 1–6. There is no clear trend visible in the doping values and the values are varying in both negative and positive carrier concentration. The GFET device yields of Runs 1–6 are 97 %, 77 %, 96 %, 99 %, 93 % and 98 %, respectively. Despite having a large back gate step height and $R_q$, Run 1 shows a high device yield, but clearly reduced electrical performance. The significant edge discontinuity in Run 2 (Figure 4b) caused a significant drop in yield (predominantly due to open circuits), which when partially resolved in Run 3 caused a significant improvement in yield.

The values obtained in Run 6 are also shown as wafer maps in Figure 6. Average $\rho_c$ (Figure 6a) in the wafer is 600 (SD = 350) Ωμm. The average $\mu_h^{TLM}$ (Figure 6c) is 2950 (SD = 170) cm$^2$/Vs. The average intrinsic doping (Figure 6b) value for all the measured GFETs in the wafer is 0.1 × 10$^{12}$ (SD = 0.4 × 10$^{12}$) cm$^{-2}$. The average hysteresis (Figure 6d) value with 20 V measurement range is 0.4 (SD = 0.9) %.

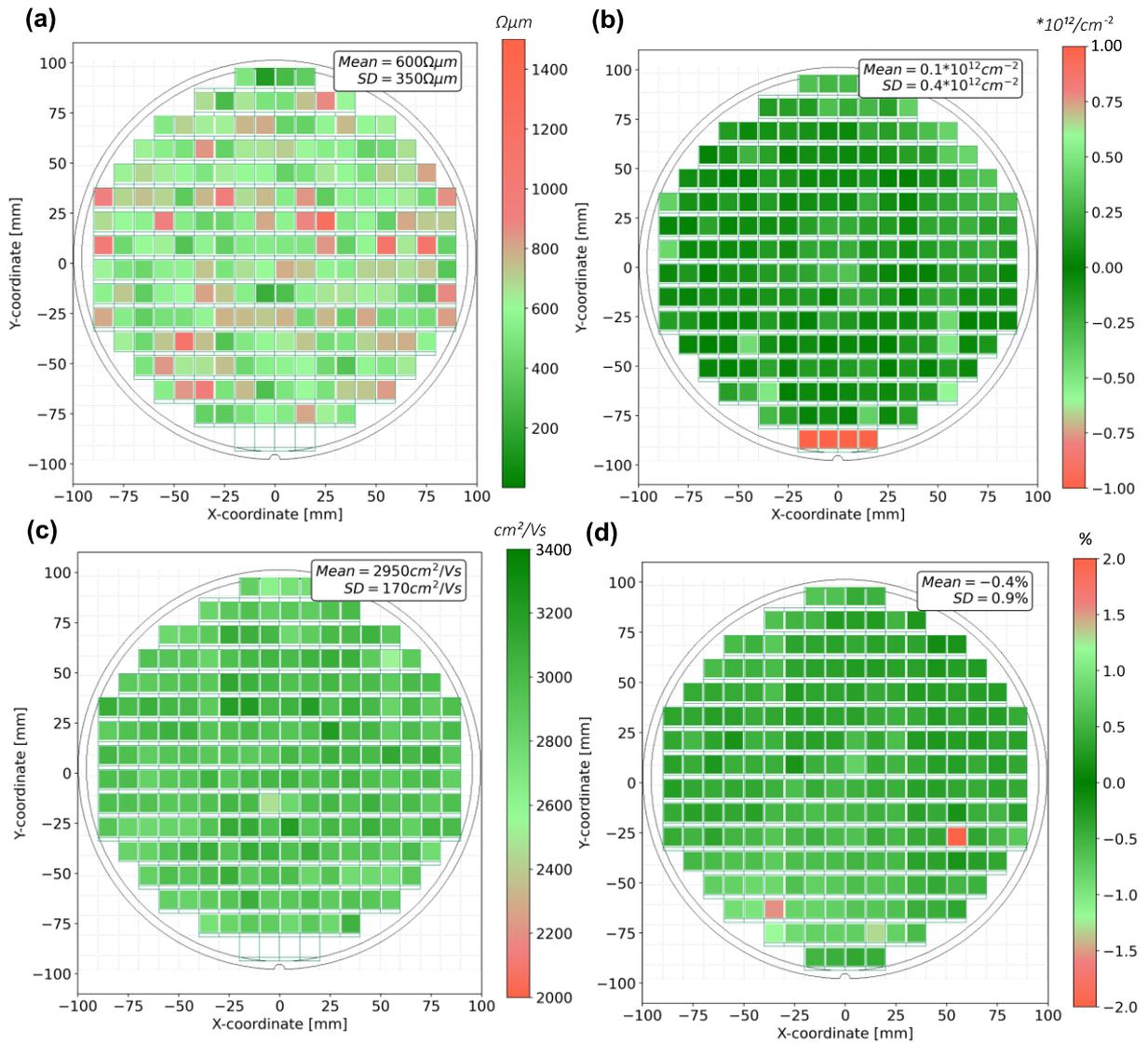

**Figure 6.** Wafer maps of process Run 6 with electrical measurements for a) specific contact resistance, b) doping, c) hole mobility and d) hysteresis. Inset values for mean and standard deviation of each parameter.

## Comparison between GFETs and gTLM structures with the gated TLM method

A comparison between GFETs and gTLM test structures was conducted in Run 6 by analysing both of these with the gated TLM method using Equations (1), (3) and (4) to extract doping, specific contact resistance, hole and electron mobility values. This is shown in Figure 7. The measurements with GFETs and gTLMs were repeated twice with six months delay after the first measurement session. 2592 gTLM lengths were measured (6 lengths per gTLM, 432 individual gTLMs total) with L = 5–30 μm and W = 10, 20 μm.

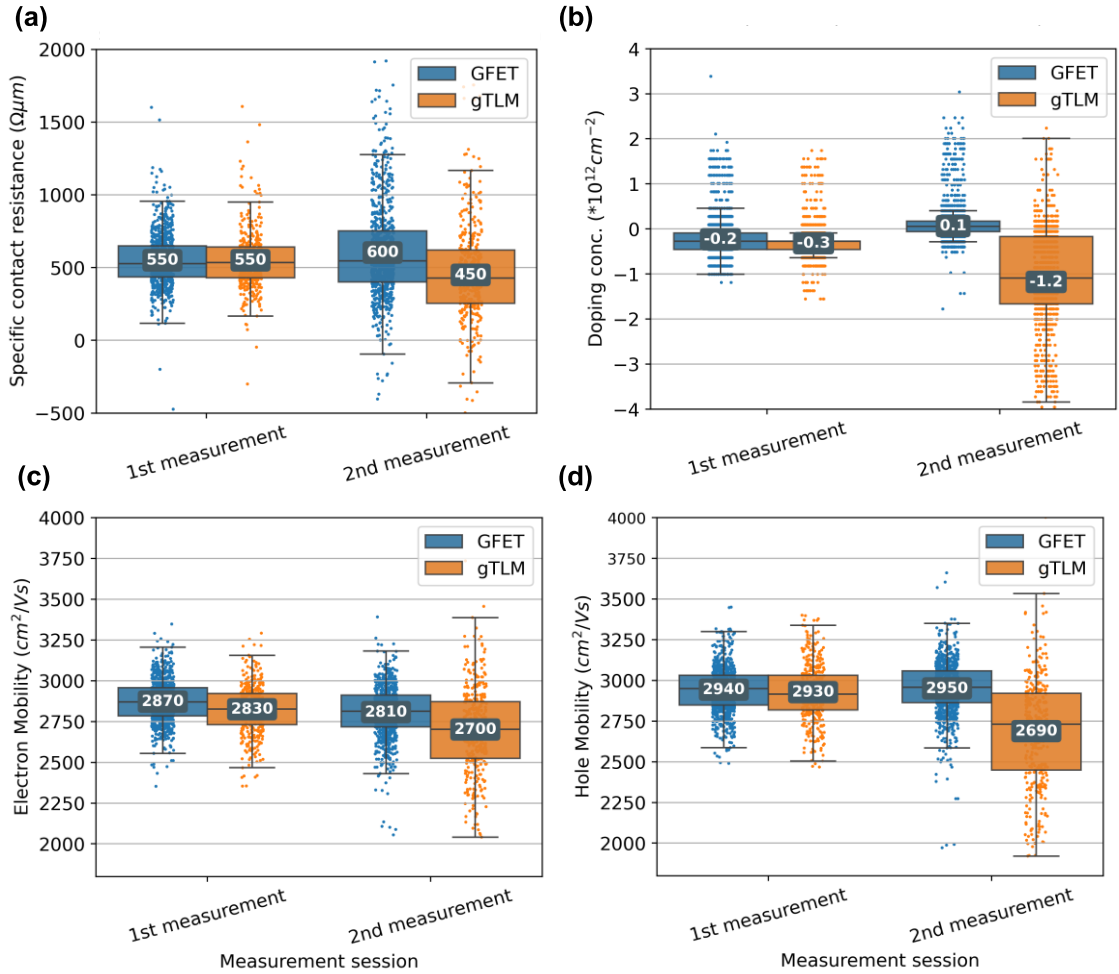

**Figure 7.** Statistical analysis of Run 6 comparing GFET and gTLM test structures with a) specific contact resistance, b) doping, c) electron mobility and d) hole mobility. GFETs are indicated in blue and gTLMs in orange. Inset values are average values of each dataset.

The average values and standard deviations for doping, specific contact resistance, hole and electron mobility are shown for the GFETs and gTLMs in Figure 7. The average doping values (Figure 7b) are -0.2 × $10^{12}$ (SD = 0.4 × $10^{12}$) $cm^{-2}$, 0.1 × $10^{12}$ (SD = 0.4 × $10^{12}$) $cm^{-2}$, for GFETs and -0.3 × $10^{12}$ (SD = 0.4 × $10^{12}$) $cm^{-2}$, -1.2 × $10^{12}$ (SD = 1.4 × $10^{12}$) $cm^{-2}$ for gTLMs across the two measurement sessions, respectively. These measurements show clearly that the GFETs give much more repeatable results with smaller variation when compared to the gTLM structures. This effect can be strongly observed from doping and also to a lesser extent from mobility. The small changes in the GFET results are likely due to time and the samples being stored in ambient conditions despite being passivated. The original condition of the samples can likely be restored by a vacuum annealing step. The larger changes and overall worse results with gTLM structures cannot be only due to the time between the measurement sessions. Another factor that can contribute to the worse gTLM results is the combined gate electrode for a set of channel lengths, as well as reused source and drain electrodes for certain channel lengths (Figure 1b). Repeated biasing on the same piece of graphene can lead to interference between the different channel lengths being measured, resulting in larger variation in device parameters. This is especially visible from the large change in doping variation in the gTLM structures in the second measurement session (Figure 7b).

**Comparison between mobility and contact resistance extraction methods**

A comparison of mobility extraction methods is shown in Figure 8a for hole mobility extracted from a series of GFETs across the wafer from Run 6. The methods shown are the gated TLM, DTM, FTM and Drude methods, and Hall measurements. In addition, we show the gated TLM method utilizing transconductance defined in Equation (6). For DTM, the hole mobility is taken from a carrier concentration of $-1 \times 10^{12}$ cm$^{-2}$ to be comparable with the other methods. Extracting the hole mobility from this carrier concentration with the DTM does not result in a significant difference in values (see Supplementary Information Figure S4). The Hall measurements were performed on individual chips from the same wafer. The total number of GFETs and gTLMs channels measured is 11232, and for Hall measurements 7 chips were randomly selected with 5 repeat measurements on each chip.

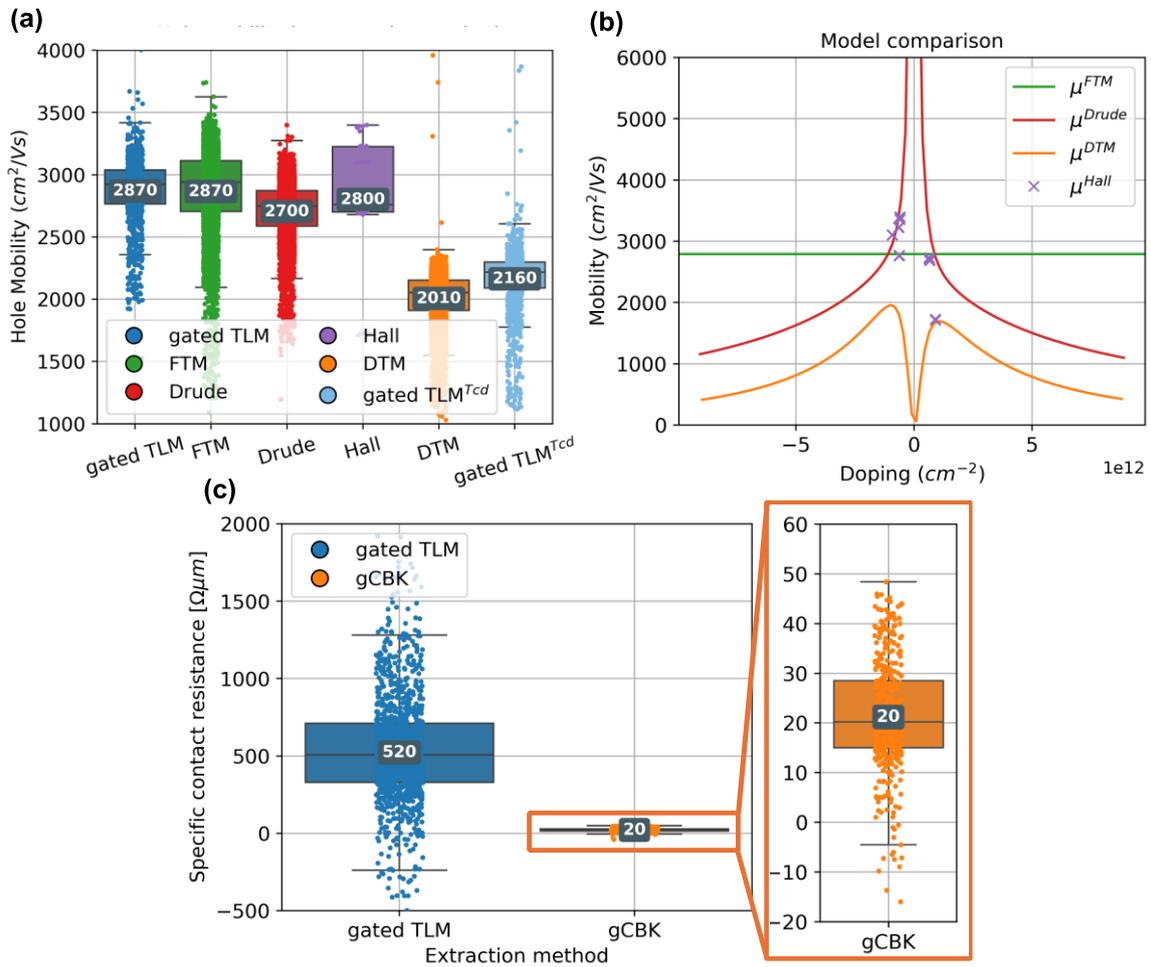

**Figure 8.** Boxplots comparing the different methods for graphene parameter extraction. a) Hole mobility from the different extraction methods. For gated TLM, Drude and DTM methods the mobility is extracted at a carrier concentration of $-1 \times 10^{12}$ cm$^{-2}$. gated TLM$^{Tcd}$ is the gated TLM method using the transconductance equation. There are 35 data points for Hall measurement (7 chips). b) Mobility of select methods against doping concentration. c) Specific contact resistance from the gated TLM method and gCBK test structures. Inset is a magnified view of the gCBK specific contact resistance. Inset values are average values of each method.

The gated TLM, FTM and Drude methods result in very similar mobility values, around 2800 cm²/Vs. These methods are different implementations of the Drude model. The gated TLM and FTM methods separate the contribution of $R_c$, which results in a higher $R_{sh}$, and thus higher $\mu_h$. The FTM additionally considers the real-nonideal characteristics of GFETs around the Dirac peak by including the $n_0$ term, which is needed to stabilize the mobility plot near $V_{dir}$. Hall measurements show average mobility of 2800 cm²/Vs, which is very close to the values obtained by the other three methods. The Hall measurements were done without gating and so the doping concentration is only measured and cannot be controlled. The average hole mobility values for the DTM and transconductance-based gated TLM method (gated TLM$^{Tcd}$) are close to each other, around 2100 cm²/Vs. The primary difference is that the DTM does not consider $R_c$. All of the methods are affected by graphene channel dimension, with larger channels producing higher mobility values, most likely due to the effect of lithographic errors being relatively greater in smaller devices.

The Drude model-based methods produce significantly higher mobility values than the transconductance-based methods (DTM and gated TLM$^{Tcd}$). The most significant difference between these two groups of methods is that transconductance (Equation (6)) defines mobility using the derivative $\partial\sigma_{sh}/\partial V_{GSO}$. This form of the equation neglects the contribution of gate induced mobility modulation in the sheet conductance derivative, and other derivatives (See Supplementary Information Section 2 for more details). Plotting the mobility extracted from these methods against doping concentration (Figure 8b) we find that the Hall measurements results support the use of the Drude model to extract mobility.

The specific contact resistance values obtained from the 5 μm × 5 μm and 10 μm × 10 μm gCBK structures in Run 6 are shown in Figure 8c and compared to $\rho_c$ from the gated TLM method. $R_c$ values calculated according to Equation (9) were normalized with the contact width using Equation (10). 432 gCBKs were measured in total. The values of $R_c$ and $\rho_c$ for the 5 μm × 5 μm and 10 μm × 10 μm CBK structures are 4 (SD = 3) Ω and 2 (SD = 1) Ω, and 20 (SD = 14) Ωμm and 22 (SD = 10) Ωμm, respectively. Both the average and standard deviation values are clearly smaller when compared to the values extracted from the GFETs or gTLMs with the gated TLM method. Despite taking into account the impact of metal lead resistances[27,44] with 4-point test structures for GFETs and gTLMs, the contact resistance values extracted with the gated TLM method by fitting show clearly higher average values and variation when compared to the values extracted with direct measurement using gCBK structures. The most likely reason for the higher values and variation obtained with the gated TLM method is due to the method heavily impacted by the sheet resistance variations between the different channels and not considering the difference in the sheet resistance in the channel and at the metal contact area.[25,32–34,42,43] Similar results with a difference between the contact resistance values extracted from gTLM and CBKs have also been obtained by Cha et al. but there is no clear explanation provided for the underlying reason.[36] In addition to the CBK structures contact end resistance (CER) test structures have been presented to overcome these problems with the traditional TLM structures.[25,32–34] FTM was also used to estimate $\rho_c$, resulting in an average value of 4460 (SD = 2470) Ωμm (not shown in Figure 8c).

**One-layer and two-layer resist approach for graphene patterning**

**AFM characterization**

The combined layer thickness of graphene and polymer residues for the Run 7 wafers were measured by AFM after the graphene patterning step for the one- (Figure 9a) and two-layer (Figure 9b) patterning approaches. For each wafer 5 locations across the wafer have been measured (Figure 9c) giving us average thicknesses of 2.5 nm (SD = 0.2 nm), 2.3 nm (SD = 0.4 nm), 1.0 nm (SD = 0.1 nm) and 1.3 nm (SD = 0.2 nm) for wafers W1, W2, W3 and W4, respectively. This indicates a clear reduction in resist residues with the utilized two-layer approach enabling cleaner graphene surfaces required for example for functionalization.

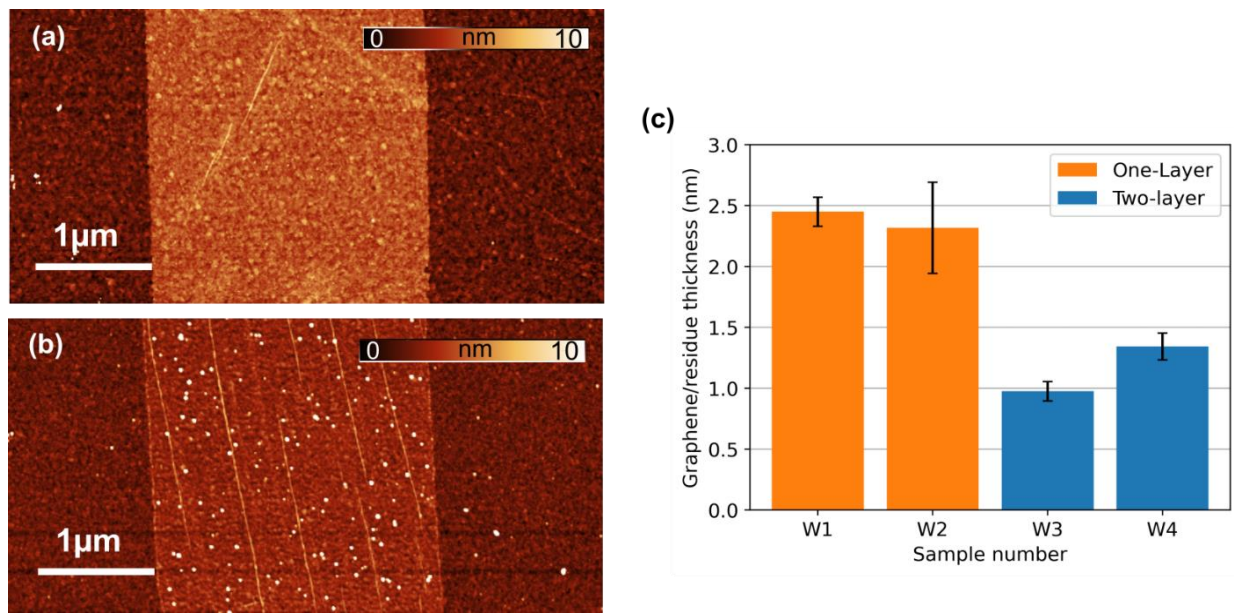

**Figure 9.** a)–b) Example AFM scans of the GFET channel area used to measure the combined thickness of graphene and resist residues. Graphene patterning was done with a) one-layer and b) two-layer resist patterning. c) The average graphene and resist thickness and standard deviation values for 4 different wafers with one- and two-layer resist patterning. One- and two-layer patterning is shown in orange and blue, respectively.

**Wafer-scale electrical characterization**

The wafers of Run 7 were electrically characterized to determine the effect of residues on the doping and uniformity of GFETs. The doping values calculated by using Equation (3) for W1 (Figure 10a) and W2 (Figure 10b) with one-layer resist patterning and for the W3 (Figure 10c) and W4 (Figure 10c) with the two-layer resist patterning show average values of 18.4 (SD = 3.8) × $10^{12}$ cm$^{-2}$, 12.9 (SD = 2.4) × $10^{12}$ cm$^{-2}$, 11.6 (SD = 1.4) × $10^{12}$ cm$^{-2}$, and 9.3 (SD = 1.2) × $10^{12}$ cm$^{-2}$, respectively. These show clear reduction with the two-layer patterning approach in overall doping as well as improvement in doping uniformity across the wafer which are very beneficial for the wafer-scale processing of GFET sensor arrays. These values have been obtained with non-passivated devices in ambient conditions which has an impact on the overall doping as well as the standard deviation values. This is not optimal for quality control devices but has been used to test the expected doping values for actual sensor arrays operated in ambient conditions. These measurements reveal that sensor characterization also in ambient conditions can provide additional insights to the process uniformity and cleanliness of the sensor surface. Comparing these results to those seen in Figure 6b, it is clear that device passivation improves doping uniformity even with the one-layer patterning approach, but also hides the non-uniformity caused by the resist residues.

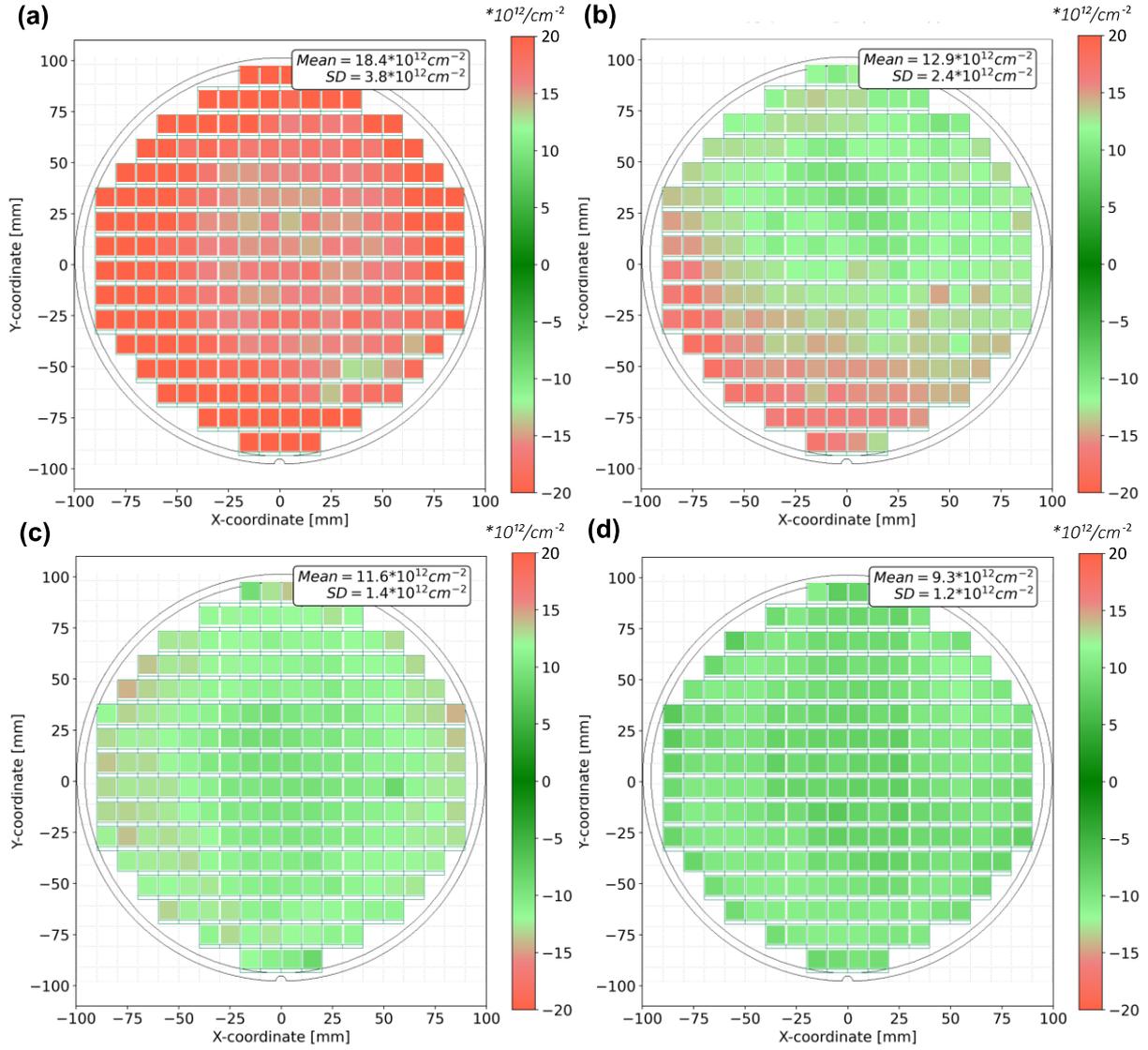

**Figure 10** Wafer maps of doping concentration measured electrically for Run 7 utilizing one-layer resist for wafers a) W1 and b) W2, and two-layer resist for wafers c) W3 and d) W4 in graphene patterning.

## 4. Conclusions

We demonstrate high yield and uniformity CMOS BEOL compatible GFET fabrication process including the gate, contact, graphene and encapsulation modules as well as statistical analysis comparing different test structures and analysis methods for the extraction of mobility, contact resistance, doping and IV-curve hysteresis values. The fabrication process and quality control protocols can be utilized for example in the wafer-scale fabrication of CMOS integrated GFET based biosensors[4], gas sensor arrays[12] and infrared cameras[11] that have been reported in the past.

The process optimization done in the process runs shows the impact of back gate roughness and back gate and contact metal step heights on the electrical performance of the devices. Overall, the roughness and metal step heights should be minimized to optimize the device performance. The comparison done between one- and two-layer resist patterned graphene channels shows a clear difference in the thickness of the polymer residue layer as well as in the overall doping and wafer-scale uniformity values. The reduction of the polymer residue layer remaining on top of the graphene after device fabrication can be critical for the subsequent graphene specific functionalization steps for example in biosensing applications.[45]

Comparison between different test structures and analysis methods also clearly shows that the selected approach can have a large impact on the values and repeatability of the measurements. GFETs analyzed with the gated TLM method show better wafer-scale uniformity for mobility, specific contact resistance and doping values when compared to gTLM structures. This is especially clear when conducting repeat measurements, with the gTLM structures showing larger variation and degradation over two measurement sessions, whereas the GFET values were much more reproducible and stable. The observed differences between the values obtained from different test structures and analysis methods need to be taken into account when comparing the values between different sources. The extracted mobility values from GFETs also depend on the selected analysis protocols as shown with the comparison between gTLM, DTM, FTM and Drude methods. This needs to be taken into account when comparing the values from different sources. Contact resistances show even higher differences between the gTLM, gCBK and FTM approaches. The lowest values are obtained with the direct measurement by using the gCBK structures while both of the fitting based approaches show higher values.

The measurement time and analysis complexity could be further simplified by using 4-point test structures such as Hall bar or Van der Pauw structures to directly measure sheet resistance for mobility extraction and using gCBK or CER structures to directly measure contact resistance. This approach would require the measurement of 2 separate test structures to extract mobility, contact resistance, doping and IV-hysteresis parameters. This would reduce the time required for the measurements when compared to the gated TLM method where minimum of 3-5 devices are typically measured.

## ASSOCIATED CONTENT

### AUTHOR INFORMATION

**Corresponding Author**

* E-mail: miika.soikkeli@vtt.fi

**Author Contributions**

§These authors contributed equally to this work.

The manuscript was written through contributions of all authors.

## ACKNOWLEDGMENT

The authors acknowledge the support by the European Union's Horizon 2020 research and innovation program under the grant agreement 2D-EPL (952792) and by the European Union's Horizon Europe research and innovation program under the grant agreements 2D-PL (101189797) and MUNASET (101119473).

## SUPPORTING INFORMATION

Capacitance measurements, comparison of different GFET models and contact resistance from cross-bridge Kelvin structures.

# REFERENCES

(1) Novoselov, K. S.; Geim, A. K.; Morozov, S. V.; Jiang, D.; Zhang, Y.; Dubonos, S. V.; Grigorieva, I. V.; Firsov, A. A. Electric Field Effect in Atomically Thin Carbon Films. *Science (1979).* **2004**, *306* (5696), 666–669. https://doi.org/10.1126/science.1102896.

(2) Li, S.; Wang, Z.; Robertz, B.; Neumaier, D.; Txoperena, O.; Maestre, A.; Zurutuza, A.; Bower, C.; Rushton, A.; Liu, Y.; Harris, C.; Bessonov, A.; Malik, S.; Allen, M.; Medina-Salazar, I.; Ryhänen, T.; Lemme, M. C. Graphene-PbS Quantum Dot Hybrid Photodetectors from 200 Mm Wafer Scale Processing. *Sci. Rep.* **2025**, *15* (1), 14706. https://doi.org/10.1038/s41598-025-96207-z.

(3) Cabral, P. D.; Domingues, T.; Machado, G.; Chicharo, A.; Cerqueira, F.; Fernandes, E.; Athayde, E.; Alpuim, P.; Borme, J. Clean-Room Lithographical Processes for the Fabrication of Graphene Biosensors. *Materials* **2020**, *13* (24), 5728. https://doi.org/10.3390/ma13245728.

(4) Soikkeli, M.; Murros, A.; Rantala, A.; Txoperena, O.; Kilpi, O.-P.; Kainlauri, M.; Sovanto, K.; Maestre, A.; Centeno, A.; Tukkiniemi, K.; Gomes Martins, D.; Zurutuza, A.; Arpiainen, S.; Prunnila, M. Wafer-Scale Graphene Field-Effect Transistor Biosensor Arrays with Monolithic CMOS Readout. *ACS Appl. Electron. Mater.* **2023**, *5* (9), 4925–4932. https://doi.org/10.1021/acsaelm.3c00706.

(5) Vieira, N. C. S.; Borme, J.; Machado, G.; Cerqueira, F.; Freitas, P. P.; Zucolotto, V.; Peres, N. M. R.; Alpuim, P. Graphene Field-Effect Transistor Array with Integrated Electrolytic Gates Scaled to 200 Mm. *Journal of Physics: Condensed Matter* **2016**, *28* (8), 085302. https://doi.org/10.1088/0953-8984/28/8/085302.

(6) Rahimi, S.; Tao, L.; Chowdhury, Sk. F.; Park, S.; Jouvray, A.; Buttress, S.; Rupesinghe, N.; Teo, K.; Akinwande, D. Toward 300 Mm Wafer-Scalable High-Performance Polycrystalline Chemical Vapor Deposited Graphene Transistors. *ACS Nano* **2014**, *8* (10), 10471–10479. https://doi.org/10.1021/nn5038493.

(7) Wu, C.; Brems, S.; Yudistira, D.; Cott, D.; Milenin, A.; Vandersmissen, K.; Maestre, A.; Centeno, A.; Zurutuza, A.; Van Campenhout, J.; Huyghebaert, C.; Van Thourhout, D.; Pantouvaki, M. Wafer-Scale Integration of Single Layer Graphene Electro-Absorption Modulators in a 300 Mm CMOS Pilot Line. *Laser Photon. Rev.* **2023**, *17* (6). https://doi.org/10.1002/lpor.202200789.

(8) Lisker, M.; Lukosius, M.; Fraschke, M.; Kitzmann, J.; Dabrowski, J.; Fursenko, O.; Kulse, P.; Schulz, K.; Krüger, A.; Drews, J.; Schulze, S.; Wolansky, D.; Schubert, A. M.; Katzer, J.; Stolarek, D.; Costina, I.; Wolff, A.; Dziallas, G.; Coccetti, F.; Mai, A. Processing and Integration of Graphene in a 200 mm Wafer Si Technology Environment. *Microelectron. Eng.* **2019**, *205*, 44–52. https://doi.org/10.1016/j.mee.2018.11.007.

(9) Canto, B.; Otto, M.; Maestre, A.; Centeno, A.; Zurutuza, A.; Robertz, B.; Reato, E.; Chmielak, B.; Stoll, S. L.; Hemmetter, A.; Schlachter, F.; Ehlert, L.; Li, S.; Neumaier, D.; Rinke, G.; Wang, Z.; Lemme, M. C. Multi-Project Wafer Runs for Electronic Graphene Devices in the European 2D-Experimental Pilot Line Project. *Nat. Commun.* **2025**, *16* (1), 1417. https://doi.org/10.1038/s41467-025-56357-0.

(10) Neumaier, D.; Pindl, S.; Lemme, M. C. Integrating Graphene into Semiconductor Fabrication Lines. *Nat. Mater.* **2019**, *18* (6), 525–529. https://doi.org/10.1038/s41563-019-0359-7.